\documentclass[runningheads,table,xcdraw]{llncs}

\usepackage[utf8]{inputenc}
\usepackage{amsmath}
\usepackage{microtype}
\usepackage{booktabs}
\usepackage{subcaption}
\usepackage{multirow}
\usepackage{tabularx}
\usepackage{hyperref}
\usepackage{todonotes}
\usepackage{wasysym}
\usepackage{array}
\usepackage{adjustbox}
\newcolumntype{x}[1]{>{\centering\arraybackslash}p{#1}}

\usepackage{xcolor}
\usepackage{tikz}

\usepackage[backend=biber,style=lncs]{biblatex}

\usepackage{todonotes}

\usepackage{xspace}

\begin{document}

\title{RemoteVote and SAFE Vote: Towards Usable End-to-End Verification for Vote-by-Mail}
\titlerunning{Towards Usable End-to-End Verification for Vote-by-Mail}
\author{Braden L. Crimmins \and Marshall Rhea \and J. Alex Halderman}
\institute{University of Michigan}

\maketitle

\vspace{-10pt}

\begin{abstract}
Postal voting is growing rapidly in the U.S., with 43\% of voters casting ballots by mail in 2020, yet until recently there has been little research about extending the protections of end-to-end verifiable (E2E-V) election schemes to vote-by-mail contexts. The first---and to date, only---framework to focus on this setting is STROBE, which has important usability limitations. In this work, we present two approaches, RemoteVote and SAFE Vote, that allow mail-in voters to benefit from E2E-V without changing the voter experience for those who choose not to participate in verification. To evaluate these systems and compare them with STROBE, we consider an expansive set of properties, including novel attributes of usability and verifiability, several of which have applicability beyond vote-by-mail contexts. We hope that our work will help catalyze further progress towards universal applicability of E2E-V for real-world elections.\looseness=-1

\end{abstract}  

\section{Introduction}
\vspace{-4pt}
In the 2020 U.S. presidential election, 43\% of voters cast their ballot by mail, twice as many as four years prior~\cite{census}. Although this increase was magnified by the COVID-19 pandemic, it follows on a long-term trend towards greater use of vote by mail~\cite{electionlab}. At least 34 states have adopted policies allowing any voter to participate by mail with no excuse, and seven of these states automatically send a mail-in ballot to every voter~\cite{ncsl20vbm}. These changes make voting by mail a fundamental part of U.S. election administration, and one that is essential to secure. Indeed, corrupting only vote-by-mail ballots would likely have allowed outcome-changing fraud in every U.S. presidential election since the turn of the century.
\looseness=-1

While end-to-end verifiable voting systems (e.g.,~\cite{helios,civitas,selene,remotegrity}) show promise for assuring election integrity, most E2E-V schemes are intended for online or in-person voting and cannot easily be adapted to secure votes cast by mail. In fact, until the recent introduction of STROBE Voting by Benaloh~\cite{strobe}, vote by mail (VbM) systems were largely absent from the verifiable voting literature.\footnote{Remotegrity~\cite{remotegrity}, which is sometimes cited as a vote-by-mail election system, relies on voters casting their ballots over the Internet after receiving them in the mail.}

We introduce two novel E2E-V systems, RemoteVote and SAFE Vote, that are designed to enable end-to-end verifiable U.S.-style postal voting while improving on aspects of usability and other desirable properties compared to STROBE\@. To evaluate our schemes and earlier VbM E2E-V proposals, we consider an expansive set of properties, including novel attributes of usability and verifiability. Several of these properties have applicability to systems outside of a VbM context. Lastly, we present a hybrid system that combines ideas from our two schemes and provides more of these properties than any other design known to the authors, although this comes at the expense of reduced practicality. We hope our work will advance the state of knowledge about E2E-V closer to systems that are suitable for deployment in real vote-by-mail elections.

\section{Related  Work: STROBE}
Benaloh's STROBE Voting~\cite{strobe} is a significant step towards achieving verifiable VbM, and identifies the primary difficulty of designing such a system: the unique limitations on communication between the voter and the election authority.

Many verifiable voting systems rely on the Benaloh challenge~\cite{benaloh-challenge,marky-usability}, a type of interactive proof that a voter can perform to check the honesty of the encrypting authority. Classically, this mechanism applies to in-person or Internet voting settings, where the voter can perform the interactive step and receive a response immediately. In VbM contexts, however, the only private communication channel between the voter and the election authority is the extremely-high-latency postal system, which imposes a difficult constraint on any kind of interactive challenge. 

Nevertheless, STROBE proposes an interactive procedure that is suitable for VbM settings. First, the election authority precomputes multiple encryptions to be sent to the voter and communicates them through the initial ballot delivery. Then, the voter chooses a set of encryptions to use and communicates this choice through the return of their voted ballot. The authority completes the challenge process by posting the decryptions of all spoiled ballots to a public bulletin board. 

This design represents a significant advance towards achieving verifiable VbM elections. Even still, as Benaloh acknowledges, there are some important practical drawbacks; in its basic form, for example, STROBE requires that each voter be sent two ballots and only be permitted to cast one. Several concerns are immediately apparent, such as the added printing and mailing costs and the potential for distrust or confusion when voters receive multiple ballots.\footnote{Considerable confusion occurred when voters were merely sent multiple vote-by-mail ballot \emph{applications} in some U.S. jurisdictions during the 2020 election cycle~\cite{abc_applications, multiple_applications}.}

Other concerns are more subtle. Suppose, for example, that a cache of ballots are returned but not processed, and the encryptions are spoiled before the mistake is discovered.\footnote{Such incidents are not uncommon, as illustrated by initially uncounted ballots uncovered during Georgia's recount following the 2020 presidential election~\cite{georgia-ap}.} Those voters lose either their privacy or their ability to validate the election outcome. Alternately, suppose both ballots in a twin set are cast. What recourse exists? This might occur due to attempted fraud by the voter, but it could equally well be the result of a misunderstanding of the system. Even if the system isn't misunderstood, it could still be an innocent mistake; for example, two members of a household might accidentally vote both ballots belonging to one of their sets and neither ballot belonging to the other. Counting both ballots would produce false evidence of fraud, because any observer could see that two twin ballots were both returned and counted. Failing to count both ballots disenfranchises one of the two voters. Neither outcome is desirable.

The STROBE paper also proposes an alternative implementation that resolves these concerns by representing two sets of encryptions on a single physical ballot, but this introduces problems of its own. For example, because the ballot is returned before the spoil process occurs, a corrupt election authority would have custody of all evidence of its malfeasance at the time it is revealed. Another concern comes from deciding which set of encryptions should be used; implicit mechanisms for choosing a column to spoil can be easily forged or manipulated by a corrupt authority, while explicit mechanisms leak information about who has participated in verification, which could facilitate targeted cheating in subsequent elections.\looseness=-1

The ideas underlying STROBE are strong, and indeed a modified version forms the basis of one of our proposals, but further advances will be required for any E2E-V system to be practically usable in real VbM elections. We hope that the two schemes we propose represent progress in that direction.
\section{Desirable Properties}
To evaluate our proposed schemes and compare them to existing designs, we introduce a set of desirable properties, ranging from attributes of the verification procedures to aspects of usability. Properties that are novel (to our knowledge) are \emph{\bfseries bolded and italicized} when first introduced.

\vspace{-4pt}
\subsection{End-to-End Verifiability}

A common goal of election system design is \textbf{end-to-end verifiability}. Systems with this property seek to provide public guarantees that the election result is accurate without relying on trust in the administering authorities. There are three main subproperties that an E2E-V system must have~\cite{Benaloh2015EndtoendV}:

\begin{description}
\setlength{\itemsep}{\smallskipamount}
    \item[Cast as Intended] means that a voter can be confident their choices were accurately received by the election authority. 
    \item[Counted as Cast] means any observer can be confident that all choices received by the election authority were tallied correctly in the final result. (This is sometimes split into \textbf{collected-as-cast} and \textbf{counted-as-collected}~\cite{bernhard2017public}.)
    \item[Eligibility Verifiability] means observers can verify that only eligible votes were included in the tally~\cite{bernhard2017public}, preventing threats like ballot-box stuffing and multiple voting. Many jurisdictions provide this property to some degree through the post-election publication of a list identifying all those who cast a ballot. This allows observers to validate that the number of votes is equal to the number of voters and (at least in principle) that all voters were eligible.%
    \footnote{Despite its importance,  this property is often ignored in the E2E-V literature, since it is primarily accomplished by existing procedural controls, rather than cryptography.}
\end{description}

When considering claims of verifiability, it is similarly important to understand who can actually perform the verification for each element of the system:

\begin{description}
\setlength{\itemsep}{\smallskipamount}
    \item[Universal Verifiability] means that any observer can single-handedly verify the property across the entire election~\cite{framework}. Typically, \emph{counted-as-cast} and \emph{eligibility verifiability} are universally verifiable in E2E-V designs.
    \item[Individual Verifiability] means that a voter can verify the correct behavior occurred for their own ballot~\cite{framework}. Typically, \emph{cast-as-intended} is individually verifiable in E2E-V designs. Mechanisms for \emph{universal cast-as-intended} exist~\cite{universal-cit}, but those are beyond the scope of this paper.
    \item[\emph{Collective Verifiability}] means that voters can collaborate with each other to verify the system's integrity. Consider a design in which a voter can gain a certain level of confidence (e.g., 50\%) that their ballot encryption is honest, but cannot individually increase their confidence beyond that level. Through collaboration with other mutually-trusting voters, each with their own independent partial confidences, these voters can jointly confirm the honesty of the system to asymptotically high degrees of collective confidence.
\end{description}

\subsection{Privacy Properties}
Another key goal of election system design is to ensure voter privacy. What precisely ``privacy'' means, however, depends heavily on the threat model, giving rise to a series of properties intended to address different threats:

\begin{description}
\setlength{\itemsep}{\smallskipamount}
    \item[Ballot Secrecy] means that a system does not leak information about voters' choices beyond that which can be deduced from the election result~\cite{bernhard2017public}.
    \item[Receipt Freeness] further means a voter cannot prove their choices to a third party after voting and casting their ballot privately\footnote{That the voter must cast their ballot \emph{privately} is implicit in most articulations of receipt freeness, which entertain only supervised voting at the polling place. Failing to articulate it in a VbM context, however, blurs the line with other stronger properties such as coercion resistance. If a voter is compelled to surrender their voted ballot and signed envelope to a coercer, this would ``prove their choices to a third party," but substantively this attack is better identified as a failure of coercion resistance.}, even if the voter actively attempts to aid in the construction of such a proof. This has historically been intended as a defense against vote buying~\cite{bernhard2017public}.
    \item[Everlasting Privacy] means that a system does not rely on assumptions about cryptographic hardness to provide the two preceding properties. If we do rely on such assumptions, a weakness in the encryption systems used---even if discovered decades later---could reveal voters' choices in previous elections~\cite{everlasting}. We further use the more nuanced terms \emph{\bfseries everlasting ballot secrecy} and \emph{\bfseries everlasting receipt freeness} to reason about systems that might only provide one of the two traits if cryptographic hardness assumptions fail.
    \item[Coercion Resistance] means the system ensures no voter can be compelled to exhibit certain voting behavior. Complete definitions entertain not only forcing a voter to choose a favored candidate, but also forcing them to abstain or to randomize their ballot, which would enable targeted disenfranchisement~\cite{bernhard2017public}.
\end{description}

\subsection{Usability Properties}
The importance of usability is well recognized in voting system design, and past failures on this front have had enormous consequences.\footnote{See, for example, the usability failures of ``butterfly ballot'' punch card systems, which some believe altered the outcome of the 2000 U.S. presidential election~\cite{chads}.} Nevertheless, voting system usability has not been formalized in the same way as other E2E-V properties. Here we attempt to define aspects of usability that can be used to discuss and compare systems (albeit with some room for subjective interpretation):

\begin{description}
\setlength{\itemsep}{\smallskipamount}
    \item[\emph{Ignorability}] means that a voter can ignore the novel elements of the system and ``vote like normal'' without compromising the integrity or privacy of their vote; they simply will not be a participant in the verification process. 
    \item[\emph{Harmlessness}] more strongly means that a voter cannot compromise the integrity or privacy of their vote even if they engage improperly with the system's novel elements. It should not be possible, for example, to inadvertently spoil a ballot by making a mistake involving the verification features.
    \item[\emph{Selection Consistency}] means that a voter's selections on their ballot will be directly reflected by the evidence that their vote was \emph{cast-as-intended}, for example in a shortcode receipt that corresponds directly to the markings they made. This property aims to prevent illusory discrepancies, like those that might arise when a ballot is processed before the evidence is generated, which might otherwise create the appearance of fraud where none exists.\footnote{For example, in a jurisdiction which permits straight-ticket voting with contest-specific overrides, the voter may expect to see a commitment to the selections they made, while a homomorphic-tallying election system would instead generate a commitment to the interpreted result of those selections after the straight-ticket rule was applied.}
\end{description}

\subsection{Error Handling Properties}
Many election systems are premised, either implicitly or explicitly, on assuming the election was error-free and designing proofs that convince observers of that fact. It is equally important, however, to entertain the possibility that errors might occur and consider how the system might handle them. The following three properties deal with identifying and addressing these errors if they occur.
\begin{description}
\setlength{\itemsep}{\smallskipamount}
    \item[Software Independence] means that an undetected error in election software or hardware cannot create an undetectable error in the election's outcome~\cite{rivest16si}. This means that no malicious software can alter the outcome undetected when there exists a dedicated and trustworthy observer.
    \item[Strong Software Independence] additionally means that when such an error is discovered, the true result can be recovered without rerunning the election~\cite{rivest16si}, for instance by hand-counting the original paper ballots.
    \item[Dispute Resolution] means that if a voter accuses an authority of dishonesty, a third party can unambiguously determine whether or not it is true~\cite{disputeresolution}. An important subproperty is \textbf{collection accountability}, which means a voter can prove it if the authority misrepresents their choice in the public tally~\cite{bernhard2017public}.
\end{description}

\subsection{Verification Process Properties}
Verification processes involve a series of checks that have properties and side effects of their own. We explicitly define these properties here, since they can have substantial implications for the reliability of the system in application:
\begin{description}
\setlength{\itemsep}{\smallskipamount}
    \item[\emph{Advance Verification}] means that all verification mechanisms can be exercised before the election results are computed. This helps to remove some improper incentives; if an individual must allege fraud before they learn the favorability of the outcome, the political motivation to make false reports is decreased.
    \item[\emph{Discretionary Verification}] means that voters are able to choose at will which elements of the system they would like to verify. This eliminates reliance on external decisional systems and allows voters to trust only their own agency. If voters are able to select arbitrary ballots to challenge for a given audit procedure, this is an example of discretionary verification. If instead the ballots are selected by some other process---for example, at random, even through some publicly accountable mechanism---the property is lost.
    \item[\emph{Independent Verification}] means that the verification mechanism can be used by an individual voter or observer without participation from the system itself. This forestalls attacks in which a cheating authority feigns inability to cooperate with the challenge process (e.g., by pretending a DDoS attack is preventing voters from logging challenge requests on a public bulletin board). It also reduces the risk due to real logistical problems during the election.
    \item[\emph{Unobservable Verification}] means that no adversary can become convinced that a given individual abstained from the verification process. If the adversary is able to learn who abstains, and they monitor this over several elections, they can target later ballot manipulation toward voters who are least likely to detect the fraud. Note that the adversary can learn the identities of \emph{some} of those who engage in the verification process, but not the identities of \emph{everyone} who does, since that would reveal exactly who does not verify.
\end{description}

\subsection{Compatibility Properties}
E2E-V schemes vary in what modes of voting and what counting methods they support. We consider two aspects of such compatibility:
\begin{description}
\setlength{\itemsep}{\smallskipamount}
    \item[\emph{Vote-by-Mail Compatibility}] means that a system can provide its intended security properties to voters whose only private communication channel with the election authority is postal mail, although such voters are assumed to also have access to publicly broadcast information.
    \item[\emph{Complexity Tolerance}] means that a system enables arbitrary computations on anonymized vote data without compromising security guarantees. Systems with this property can implement any computable tallying system, such as instant-runoff voting. Implementations without this property, including homomorphic systems, are limited in what voting methods they can support.
\end{description}

\section{Our Proposed Schemes}

\vspace{-4pt}
Our schemes use a similar approach to STROBE~\cite{strobe}. At a high level, every voter is given a ballot with a unique ID and seemingly random shortcodes beside each candidate. When voting, they can record this ID and the shortcodes corresponding to their choices. The authority posts this same information to a public bulletin board for every ballot they collect, and voters can confirm this ``receipt'' matches what they recorded to be confident their votes were counted correctly.

There are similarities behind the scenes, too. As with STROBE, we use ElGamal threshold encoding to create a homomorphic encryption for each choice available to the voter.\footnote{Like STROBE, we include abstentions and undervotes within the meaning of ``choice.'' For this reason, in a contest where the voter can select $k$ candidates, we produce $k$ encryptions representing abstentions in addition to the ones representing selected candidates. This way, all choices---including abstentions and undervotes---can be represented as the sum of $k$ encryptions, and are therefore indistinguishable.} We also generate a shortcode for each candidate that commits to the encryption, such that the mapping between codes and candidates is only known to the voter but observers can use the published shortcode receipt to verify the election tally. Lastly, we generate a single longcode that commits to the encryptions without revealing the candidates they correspond to.

\begin{figure}[t]
    \centering
    \begin{subfigure}{0.465\linewidth}
        \centering
        \includegraphics[width=\linewidth]{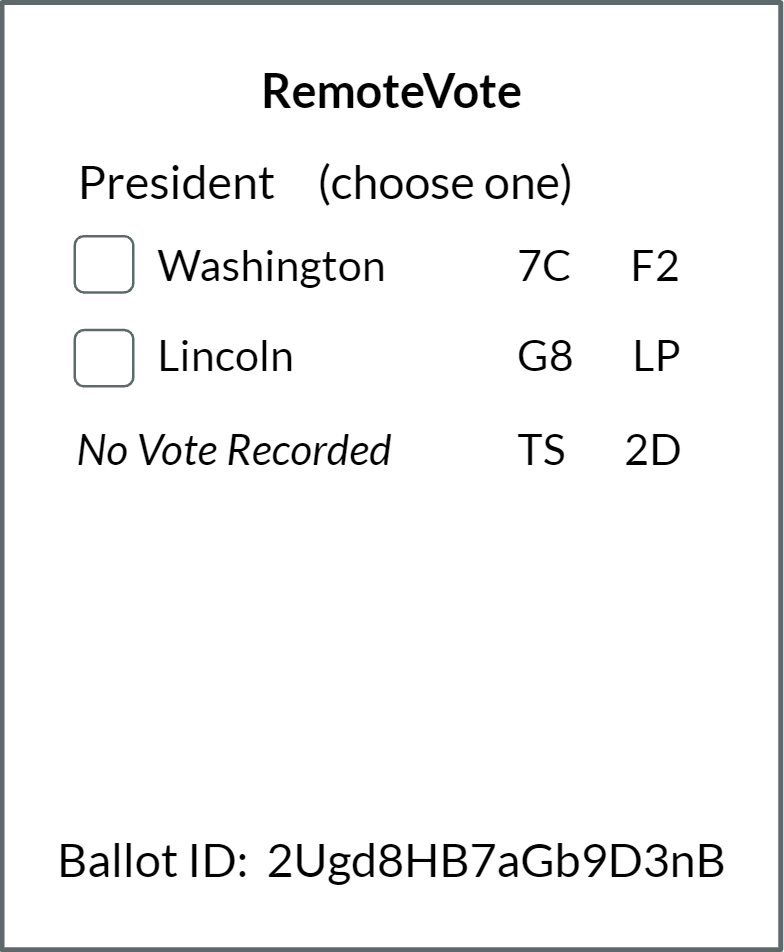}
        \caption{RemoteVote}\vspace{-4pt}
        \label{subfig:remotevote-ballot}
    \end{subfigure}\quad
    \begin{subfigure}{0.465\linewidth}
        \centering
        \includegraphics[width=\linewidth]{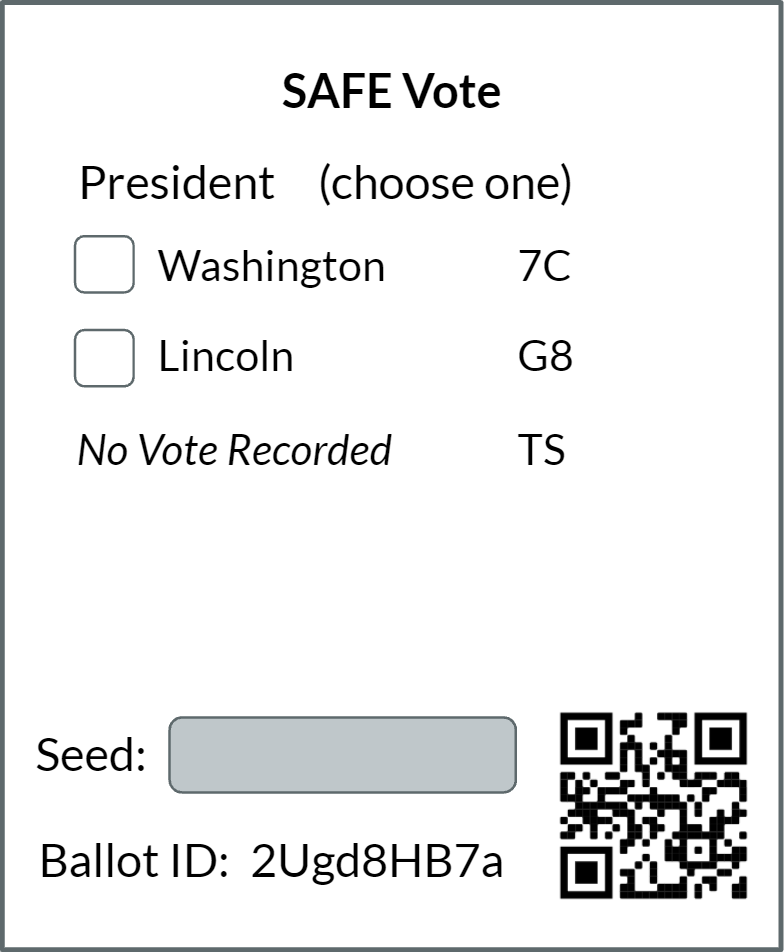}
        \caption{SAFE Vote}\vspace{-4pt}
        \label{subfig:safevote-ballot}
    \end{subfigure}
    \caption{Ballot designs for our two proposed schemes.}\vspace*{-12pt}
    \label{fig:ballots}
\end{figure}

We also introduce a series of departures and enhancements compared to the STROBE framework. First, when we tally votes for a contest, we do it in a two-stage process. We begin by homomorphically tallying each individual voter's choices within the contest to generate a single encryption per ballot cast. We then run these encryptions through a verifiable anonymizing mixnet (e.g.,~\cite{shuffling,mixnet}) then decrypt and process them in the clear. This means we can represent candidates with a single binary flag each, denoting either that they were selected or were not. Also, there is no need for STROBE's non-interactive zero knowledge proofs that each encryption contains a valid vote; any issues will be recognized when the decryption process occurs, and can affect at most a single voter. The ordinary challenge process suffices to address this risk. A final advantage is that this approach enables \emph{selection consistency} and \emph{complexity tolerance}; because we process the data in the clear, we support arbitrary computation on voters' anonymized choices and therefore support arbitrary tallying mechanisms. STROBE, by contrast, relies on homomorphism to compute the result, which means it cannot support voting systems that use more complicated tallying methods and cannot always directly represent a voter's selections in the published shortcode receipts.
\looseness=-1

Second, when we generate our encryptions, we use a random value $R$ to seed a PRNG, and use the output to seed ElGamal for every encryption on a given ballot. This way, revealing $R$ to a voter allows them to reconstruct all the encryptions on their ballot, a property used in SAFE Vote's challenge process.

Third, we do not directly reveal the encryptions involved in our system to the public. Instead, we derive perfectly hiding commitments to these encryptions via the PPATS mechanism introduced in~\cite{cce} and further elaborated on in~\cite{everlasting}. This system is an example of a \emph{commitment consistent encryption} (CCE) scheme, which allows the creation of commitments that can be used as a one-to-one stand-in for encryptions in a homomorphic tally or mixnet, then can be ``opened'' to the underlying encryption value once they are tallied or anonymized. These commitments offer information theoretic privacy~\cite{everlasting}, which means the privacy properties remain even if all modern cryptographic assumptions fail. We generate these commitments immediately after generating our encryptions and use them---rather than the encryptions themselves---to derive shortcodes and ballot IDs. After the mixnet step in our tallying, we open the commitments to the underlying encryptions before decrypting. The process is otherwise as previously described.\footnote{Using CCEs to implement both a homomorphic and mixnet tallying step is usually infeasible for large-scale elections~\cite{cce}. However, because the format of possible votes is highly predictable, an exhaustive search of plaintexts is easy, and decryption becomes cheap enough that we can use the PPATS system for both.} Note that each commitment requires 256 bits of true randomness to generate, which is retained privately by the election authority for use in the tallying process.

Fourth, and lastly, we commit to each encrypted ballot in advance of the election. This means posting the commitments, shortcodes, and ballot ID of each ballot when it is first generated.\footnote{Alternately, we can simply post the commitments (which can be used to derive the rest of the information), but posting everything makes the process easier for observers.\looseness=-1} This is not strictly a difference from STROBE, as this early commitment is still possible in the original design. It is essential for RemoteVote's challenge process, however, and so is important to specify here.

\subsection{RemoteVote}
RemoteVote ballots (Figure~\ref{subfig:remotevote-ballot}) look identical to those of single-ballot STROBE, but the challenge process is markedly different. While STROBE attempts to facilitate an interactive challenge through limited communication channels, RemoteVote abandons the idea of interactivity altogether.

This means we sacrifice \emph{discretionary auditing}, but in exchange we gain a number of other valuable properties. Notably, we can perform the challenge process when voters still have custody of their ballots, giving a greater degree of \emph{dispute resolution}. There is also no way for the authority to be confident that any voter abstained from this process, giving \emph{unobservable verifiability}. Recall too that in single-ballot STROBE, the election authority could simply ignore the preference of the voter and spoil whichever column it wanted. In our design, this is no longer the case; compliance with the spoiling procedure is \emph{universally verifiable}.
\looseness=-1

To implement this scheme, we begin by pairing encrypted ballots.\footnote{Pairs should be constructed such that there is no shortcode collision within a contest.} We commit to these pairings by posting them to a public bulletin board and generating a single combined ID from their two longcodes. The paired set is printed on a single physical ballot such that there are two columns of shortcodes beside the candidates' names, one column corresponding to each encrypted ballot. The ballot ID is printed at the bottom of each ballot in human-readable form.

After the ballots are delivered to voters, randomness is generated in a publicly accountable manner (e.g.,~\cite{rla}) and used to deterministically select one column to spoil for each ballot.\footnote{One effective heuristic would be hashing the randomness with the ballot ID and selecting a column based on the resulting hash’s parity.} The secret information for that column’s encrypted ballot is posted publicly alongside all the rest of the ballot pair’s public information, where it can be accessed at will by third parties. These observers can then validate that the column's encryptions and commitments were all well formed.

With the newly public information, a voter can apply a trusted third-party system to generate a partial image for their ballot based on its ID\@. This image would display one column of expected shortcodes next to the appropriate candidates, and by validating this against the ballot they were sent, the voter can be sure that the spoiled column of shortcodes represented an accurate set of encryptions. Provided the voter trusts the public randomness, they know that a forged encryption on their ballot would have had a 50\% chance of being revealed.\footnote{Even if the voter does not trust the \emph{entire} public randomness, they can still have confidence in this system. Only one in $2^{2^n}$ possible pairings would be fully resistant to detection, where $n$ is the number of trustworthy bits. Validating that a particular pairing belongs to that set would require computing $2^n$ hashes if the challenge heuristic uses a secure hash function. This means that an attack on encryption integrity is computationally infeasible for any meaningful number of trustworthy random bits.} This provides \emph{collective verifiability}, because many audits at 50\% confidence quickly scale to ensure any large-scale fraud would be detected.

The voter can then vote like normal (\emph{ignorability} and \emph{harmlessness}), return their ballot, and optionally watch that one of the two shortcodes beside each selected candidate appears for the proper contest. Whether that shortcode is from the correct column can be validated by anyone else, because the spoiling of the challenged column is \emph{universally verifiable}.

\subsection{SAFE Vote}
Scratch Auditing for Fair Elections (SAFE) Vote is an alternate approach that attempts to preserve the interactivity of the challenge process while resolving some of STROBE's other issues. We accomplish this by including a physical challenge mechanism on each ballot. This maintains \emph{discretionary auditing}, but also incorporates \emph{independent auditing} and ensures that the challenge process provides immediate evidence as to whether or not a discrepancy was uncovered. Like RemoteVote, this occurs when the ballot is still in the voter's custody, giving \emph{dispute resolution} through physical evidence of potential fraud.

SAFE Vote ballots (Figure~\ref{subfig:safevote-ballot}) contain four special features:
\begin{enumerate}
    \item Each option has a single corresponding shortcode printed beside it.
    \item The longcode is printed as the ballot ID\@.
    \item The value of $R$ is printed on the ballot and concealed by a scratch-off surface.
    \item The ballot includes a large QR code, which, when decrypted using $R$ as the key, provides the true randomness used to generate the ballot's commitments.\footnote{Recall that 256 bits of true randomness are necessary per commitment.}
\end{enumerate}

These modifications can be ignored by voters who do not choose to participate in the verification process, but they enable voters to independently audit the ballot's encryptions if they desire. To do so, voters would remove the scratch-off surface, revealing $R$. Trusted third-party software can use this $R$ value along with the QR code to reproduce the ballot’s shortcodes and ID in full, and the voter can compare the output against the ballot they were sent. If there are no inconsistencies, voters can be confident that their ballot was correctly encrypted. They can then request a replacement and repeat the challenge as many times as they like, until they are satisfied with the encryptions and cast a ballot.\footnote{Voters may also simply request multiple ballots up front, if policy permits.}

Ballots received in this system should be processed just as with STROBE, with one exception: in the event that a voter returns a ballot with a removed scratch-off surface, no shortcode receipt should be published. Providing such a receipt when $R$ had been revealed would violate the property of \emph{receipt freeness}, since the two pieces of information combined would reveal the choices of the voter. Instead, the election authority should make a post to the bulletin board identifying that the ballot had its scratch-off removed. The voter’s choices should then be transcribed faithfully onto a new encrypted ballot by a multipartisan ballot duplication team, like those currently employed in absent voter counting boards across the country. This new ballot can be cast as normal, so that the voter’s selections still appear in the public tally. This ensures that the addition of the system’s novel properties cannot introduce new sources of disenfranchising voter error, thereby providing the property of \emph{harmlessness} to our scheme.\footnote{This also introduces the risk that an adversary might change the voter's selections and remove the scratch-off surface to eliminate any direct evidence that the change occurred. A potential solution might be to introduce a grace period where voters who notice their ballots were duplicated in this process can instead spoil them and revote.}
\renewcommand{\topfraction}{.75}
\renewcommand{\textfraction}{.1}
\def\defeq{\mathrel{\mathop:}=}

\begin{figure}
    \centering
    \begin{minipage}[t]{0.52\textwidth}
    \raggedright
    \scriptsize
    
    \textbf{Initial Preparation (both schemes):}\vspace{-6pt}
    \begin{enumerate}
    \item Generate election paramters and vote data
    \item For each ballot:
        \begin{enumerate}
	    \item Roll random $R$ to seed ElGamal
	    \item Encrypt each piece of vote data
	    \item Gen.\ CCE com'ts to the encryptions
	    \item Gen.\ shortcodes/longcode from com'ts
	    \end{enumerate}
    \end{enumerate}
    \vspace{-3pt}
	
	\textbf{RemoteVote Preparation:}\vspace{-6pt}
    \begin{enumerate}
    \item Pair ballots, avoiding shortcode collisions
    \item For each pair:
        \begin{enumerate}
            \item Compute $\text{ballot ID}\defeq\text{Hash}(\text{longcodes})$
            \item Publish ID, longcodes, com'ts, shortcodes\looseness=-1
            \item Print ballot (2 shortcode cols., ballot ID)\looseness=-1
	    \end{enumerate}
	\item Send ballots to voters
    \end{enumerate}
    \vspace{-3pt}
        
    \textbf{RemoteVote Audit Process:}\vspace{-6pt}
    \begin{enumerate}
        \item After voters receive ballots, $R_p \defeq\text{pub.\ randomness}$
        \item For each ballot, $\text{Hash}(R_p \parallel\text{ballot ID})$:
        \begin{itemize}
            \item If even, publish secret data for col.\ 1
            \item If odd, publish secret data for col.\ 2
        \end{itemize}
	   \item Observers can then:
	   \begin{enumerate}
	       \item determine if correct column was spoiled 
	       \item associate spoiled shortcodes with cands
	   \end{enumerate}
    \item Voter compares association with their ballot.\\If discrepancy, raise error; else, cast normally
\end{enumerate}
	
    \end{minipage}\begin{minipage}[t]{0.48\textwidth}
    \scriptsize
    
    \textbf{SAFE Vote Preparation:}\vspace{-6pt}
    \begin{enumerate}
    \item For each ballot:
	    \begin{enumerate}
	    \item Publish longcode, com'ts, shortcodes
	    \item Print ballot $($column of shortcodes,\\
        \strut\quad longcode printed as ballot ID,\\
		\strut\quad random $R$ beneath scratch-off,\\
        \strut\quad $\text{QR code}\defeq \text{Enc}(R, \text{commit rand.}))$
	    \end{enumerate}
	\item Send ballots to voters
	\end{enumerate}
	\vspace{1pt}

	\textbf{SAFE Vote Audit Process:}\vspace{-6pt}
\begin{enumerate}
    \item Voter optionally reveals scratch off, then:
    \begin{enumerate}
        \item Uses $R$ and QR code to generate full ballot image
	    \item Compares with printed shortcodes and ballot ID. If okay, encryptions honest
	    \item Requests replacement ballot, repeats
	\end{enumerate}
	\item Else, voter casts ballot as normal
\end{enumerate}
\vspace{1pt}

\textbf{Ballot Return/Counting (both schemes):}
\vspace{-6pt}
\begin{enumerate}
\item Authority receives/validates ballot, then\\publishes ballot ID and selected shortcodes
\item Voter optionally compares w/ their choices
\item Homomorphically tally choices by contest
\item Anonymize aggregated com'ts via mixnet
\item Open com'ts, decrypt, and process in clear
\item Observers validate announced result
\end{enumerate}    

    \end{minipage}
    \caption{Procedures for initial setup, verification, and counting in our two schemes.}
    \label{fig:pseudocode}
\end{figure}

\section{Extensions}

\subsection{Collection Accountability}
So far, our schemes lack \emph{collection accountability}, which is vital for \emph{dispute resolution}. It requires that voters be able to prove which commitments they selected, so a corrupt authority cannot misrepresent their choices on the bulletin board.
\looseness=-1

We can extend our schemes to accomplish this by allowing voters who select a candidate to learn a portion of the corresponding encryption through some mechanism like Remotegrity's scratch-off surfaces~\cite{remotegrity}. Then, if the election authority posts the wrong shortcodes to the bulletin board, the voter can identify the commitments they actually selected and provide the information about the secret encryptions underlying those commitments as evidence of their choice.\looseness=-1

Because no observers are able to verify that the partial encryptions the voter presents are legitimate, this proof must be interactive. After the voter presents the information, the authority can open the commitment to the true underlying encryption---which should be inconsistent with the voter's claim---and present a non-interactive zero-knowledge proof that the opened encryption is a valid vote. If the authority is unable to do so, the challenge is presumably valid.

Note that the strength of this proof depends on the integrity of the hiding mechanism. If the encryption for a given candidate became known to an attacker improperly---say by using advanced imaging technology, compromising the printer, or simply by revealing all the information on a ballot and then counterfeiting a new one where the information was still hidden---they could use it to wrongfully implicate the authority for malfeasance. This increases the difficulty of making false allegations, but it may make them more persuasive when well executed.

There are a number of trade-offs associated with this modification. For one, we would lose \emph{everlasting receipt freeness}, because a voter could make a false challenge claiming they selected a commitment they really did not. The election authority would then open this commitment to disprove the claim, revealing the encryption underneath. This allows the voter to produce a receipt that would reveal a candidate they did \emph{not} select once cryptographic assumptions fail.\footnote{If an attacker can impersonate a voter when making a challenge, they could use a similar strategy to compromise \emph{everlasting ballot secrecy} by making false challenges that reveal other voters' unchosen encryptions.}

Additionally, if a voter made a selection and then changed their mind, they would have evidence they could use to falsely implicate an honest authority, but forbidding them from undoing a marking loses \emph{harmlessness}. We can solve this by allowing the authority to post a disclaimer for such ballots, noting that a contrary proof might appear. A corrupt authority could abuse this to evade collection accountability, but it would place a bound on the number of cheated ballots, and---as with SAFE Vote---we can give a grace period for the voter to spoil and recast their ballot if it has a reported issue.

Note too that collection accountability guarantees only apply to active misrepresentation in the public tally. Ballots could still be susceptible to being excluded from the tally by an attacker who intercepts and discards them, and there is no apparent cryptographic mechanism to guard against this in a VbM context.
\vspace{-8pt}

\subsection{Assigning Voters Private Keys}
Another possible extension would be to generate a key pair for each ballot. The public key would be posted to the bulletin board alongside the rest of the ballot's information, and the private key would be printed on the ballot to be disclosed directly to the voter. This would allow the voter to sign their challenge for the collection accountability extension, eliminating the threat to \emph{everlasting ballot secrecy}. This would also allow for procedural controls that keep the authority from learning the mapping between voters and ballots while still allowing voters to exercise the revoting mechanisms discussed previously.\footnote{Note that this breaks \emph{harmlessness}; a voter who leaks their private key and makes a mistake that would allow revoting would compromise the integrity of their ballot.}
\vspace{-8pt}

\subsection{System Synthesis}
RemoteVote and SAFE Vote each have their own advantages and limitations, but by combining them we can get the best of both worlds, albeit in a less practical form. In this hybrid system, we print each ballot with two columns of shortcodes, one of which is spoiled through RemoteVote's public audit process, and the other of which can be spoiled at will by removing a SAFE Vote style scratch-off surface.

\newcommand{\Y}{$\CIRCLE$}
\newcommand{\N}{\Circle}
\newcommand{\I}{\LEFTcircle}
\begin{table}[t]
    \small
    \centering
    \begin{tabular}{rx{1.2cm}x{1.2cm}x{1.2cm}x{1.2cm}x{1.2cm}}
    \toprule
    &
    \rotatebox{90}{RemoteVote} &
    \rotatebox{90}{SAFE Vote} &
    \rotatebox{90}{Synthesized} &
    \rotatebox{90}{STROBE} &
    \rotatebox{90}{Single-ballot} \rotatebox{90}{STROBE} \\
    \midrule
    Individual Cast-as-Intended & \N & \Y & \Y & \N & \N \\
    Collective Cast-as-Intended & \Y & \Y & \Y & \Y & \Y \\
    Counted-as-Cast & \Y & \Y & \Y & \Y & \Y \\
    Eligibility Verification & \Y & \Y & \Y & \Y & \Y \\
    \cmidrule{1-6}
    Ballot Secrecy & \Y & \Y & \Y & \Y & \Y \\
    Receipt Freeness & \Y & \Y & \Y & \Y & \Y \\
    Everlasting Ballot Secrecy & \Y & \Y & \Y & \I & \I \\
    Everlasting Receipt Freeness & \I & \I & \I & \I & \I \\
    Coercion Resistance & \N & \N & \N & \N & \N \\
    \cmidrule{1-6}
    Ignorability & \Y & \Y & \Y & \N & \Y \\
    Harmlessness & \I & \I & \I & \N & \Y \\
    Selection Consistency & \Y & \Y & \Y & \N & \N \\
    \cmidrule{1-6}
    Software Independence & \Y & \Y & \Y & \Y & \Y \\
    Strong Software Independence & \Y & \Y & \Y & \Y & \Y \\
    Collection Accountability & \I & \I & \I & \N & \N \\
    Dispute Resolution & \I & \I & \I & \N & \N \\
    \cmidrule{1-6}
    Advance Verification & \Y & \Y & \Y & \Y & \Y \\
    Discretionary Verification & \N & \Y & \Y & \Y & \I \\
    Independent Verification & \N & \Y & \Y & \N & \N \\
    Unobservable Verification & \Y & \N & \Y & \Y & \I \\
        \cmidrule{1-6}
    Vote-by-Mail Compatibility & \Y & \Y & \Y & \Y & \Y \\
    Complexity Tolerance & \Y & \Y & \Y & \N & \N \\
    \bottomrule
    
    \multicolumn{6}{c}{\Y$=$provides property\quad\N$=$doesn't provide property}\\
    \multicolumn{6}{c}{\I$=$property is implementation dependent}
    \end{tabular}
    \medskip
    \caption{Comparison of properties achieved by our schemes and by STROBE\@.}
    \label{table:properties}
\end{table}

This allows for the unification of a number of advantages. Voters can get \emph{discretionary verification} and \emph{individual verification} by choosing to remove the scratch-off and spoil their ballot as they do under SAFE Vote, but a corrupt authority still has no way of reliably identifying safe targets for cheating; voters who do not remove the scratch-off still may have verified the publicly spoiled column, giving \emph{unobservable auditing} as well. We also gain \emph{advance auditing}, because all challenges can be performed before results are published.

The disadvantage is that this system has higher complexity. Between the two columns of shortcodes, two encryptions hidden beside every candidate, two $R$ values concealed under scratch-offs at the bottom, a secret key used to sign challenges, and 512 bits of physically printed randomness per candidate (likely in the form of multiple large QR codes), the ballot would be challenging to use for those wishing to participate in the verification process. They would need to track and make sense of all the novel information presented to them. Perhaps future implementations can maintain these properties while simplifying the mechanism.
\section{Comparisons and Conclusion}

Our schemes show that E2E-V can be used to secure VbM with practical usability. They build heavily on STROBE~\cite{strobe}, the first VbM-compatible E2E-V framework.
\looseness=-1

We compare the detailed properties of our schemes to STROBE in Table~\ref{table:properties}. As shown there, RemoteVote expands on the guarantees provided by single-ballot STROBE, while SAFE Vote trades the property of unobservable verification to gain individual cast-as-intended verification, along with discretionary and independent verification. The synthesized approach we introduced in Section 5.3, meanwhile, provides the strongest set of guarantees of any of these systems.

Other mechanisms for comparison exist as well; the formal properties are merely starting points. For example, STROBE fails to resolve disputes where the authority misrepresents the voter's choices in the published shortcode receipts, while single-ballot STROBE \emph{additionally} loses dispute resolution for allegations that a ballot's shortcodes were printed beside the wrong candidates. RemoteVote and SAFE Vote are able to solve these problems with appropriate extensions, but they are still vulnerable to an attack against the processing procedure instead of ballot integrity; e.g., an adversary could unaccountably intercept and discard ballots. All this nuance is lost when dispute resolution is regarded as binary.

Much remains to be done, including producing rigorous proofs of these schemes' properties, creating and field-testing prototypes, and potentially further simplifying the voter experience. Nevertheless, we believe RemoteVote and SAFE Vote are important steps toward usable verifiability for vote-by-mail, which is a necessary precondition for the universal adoption of E2E-V in U.S. elections.

\vspace{-3pt}
\subsection*{Acknowledgements}

We thank Josh Benaloh and Olivier Pereira for insightful discussions and feedback. This work was supported by the Andrew Carnegie Fellowship, the U.S. National Science Foundation under grant no.\ CNS-1518888, and a gift from Microsoft.
\vspace{-3pt}

\printbibliography

\end{document}